\documentstyle[12pt]{article}

\newcommand{\be}{\begin{equation}}
\newcommand{\ee}{\end{equation}}
\newcommand{\ba}{\begin{eqnarray}}
\newcommand{\ea}{\end{eqnarray}}
\newcommand{\non}{\nonumber\\ }

\newcommand{\Gp}{\Gamma_{(p)}}
\newcommand{\pS}{\partial\Sigma}
\newcommand{\umu}{\underline{\mu}}
\newcommand{\unu}{\underline{\nu}}

\newcommand{\bmu}{\overline{\mu}}
\newcommand{\bnu}{\overline{\nu}}

\begin{document}

\renewcommand{\thefootnote}{\fnsymbol{footnote}}
\font\csc=cmcsc10 scaled\magstep1
{\baselineskip=14pt
 \rightline{
 \vbox{\hbox{UT-807}
       \hbox{YITP-98-15}
       \hbox{February 1998}
}}}

\vfill
\begin{center}
{\large\bf
Matrix Model for Dirichlet Open String}

\vfill

{\csc Kiyoshi EZAWA}$^{1)}$\footnote{JSPS fellow.
      e-mail address : ezawa@yukawa.kyoto-u.ac.jp},
{\csc Yutaka MATSUO}$^{2)}$\footnote{
      e-mail address : matsuo@hep-th.phys.s.u-tokyo.ac.jp},
{\csc Koichi MURAKAMI}$^{1)}$\footnote{
      e-mail address : murakami@yukawa.kyoto-u.ac.jp}\\

\vskip.2in

$^{1)}${\baselineskip=15pt
\it  Yukawa Institute for Theoretical Physics, \\
  Kyoto University, Sakyo-ku, Kyoto 606-8502, Japan \\
}

\vskip.2in

$^{2)}${\baselineskip=15pt
\it Department of Physics, Faculty of Science, University of Tokyo,\\
     Hongo, Bunkyo-ku, Tokyo 113, Japan\\
}
\end{center}

\vfill
\setcounter{footnote}{0}
\renewcommand{\thefootnote}{\arabic{footnote}}

\begin{abstract}
We discuss the open string ending on D $p$-branes
in IKKT framework.
First we determine the boundary conditions
of Green-Schwarz superstring
which are consistent with supersymmetry and $\kappa$-symmetry.
We point out some subtleties arising from
taking the Schild gauge and show that in this gauge
the system incorporates
the limited dimensional D $p$-branes ($p=3,7$).
The matrix regularization for the Dirichlet open string
is given by gauge group $SO(N)$.
When $p=3$, the matrix model becomes the dimensional reduction
of a 6 dimensional ${\cal N}=1$ super Yang-Mills theory.

\end{abstract}

hep-th/9802164
\setcounter{footnote}{0}
\renewcommand{\thefootnote}{\arabic{footnote}}
\vfill
\newpage

\section{Introduction}                                  

During the recent developments in the superstring theory,
the matrix theories give a new framework
to understand the non-perturbative aspects of the string
theory\cite{r:bfss}. 

Among some variants of such theories,
we focus on IKKT matrix model \cite{r:ikkt} 
which is the large-$N$ reduced model of ten dimensional
$U(N)$ super Yang-Mills theory.
It is supposed to give a constructive definition of the
type IIB superstring theory in the Schild gauge.
To understand the non-perturbative properties,
it is indispensable to have an appropriate description
of $p$-branes.  Originally, such objects were studied
from classical configuration with $[X^i,X^j]\propto {\bf 1}$
\cite{r:bfss,r:ikkt,r:bps}.


In this paper, we propose another
picture of $p$-branes by defining the open string
ending on D $p$-branes, so-called {\it Dirichlet open string},
in IKKT framework. It is an extension of our
previous work \cite{r:EMM3} where open supermembrane
ending on $M$ five-brane was investigated by
using the regularization technique similar to
that in \cite{r:dWHN}.

In sec.\ref{sec:bciib}
we determine the boundary conditions for the 
type IIB Green-Schwarz open superstring
which are consistent with space-time supersymmetry
and $\kappa$-symmetry.
We describe the boundary conditions for fermions
in a linear form and discuss subtleties
in the analytic continuation
of fermionic variables which is mandatory for IKKT model.

In sec.\ref{sec:mros} we account for implications
coming from the Schild gauge.  One of the main consequences is
that the system is consistently accompanied only by
the limited dimensional D $p$-branes $(p=3,7)$.
We then identify an unbroken part of supersymmetry.
Finally we show that such an open string is
regularized by $SO(N)$ matrix model.
When $p=3$, in particular, it becomes the reduced model of a
6 dimensional ${\cal N}=1$ super Yang-Mills theory.

In appendix we summarize the convention of
$SO(1,9)$ gamma matrices that we use in this paper.

\section{Boundary Conditions for the Type IIB Superstring} 
\label{sec:bciib}                                         
In this section,
we fix the boundary conditions of the
type IIB Green-Schwarz (GS) superstring ending on D $p$-branes. 
We consider the case in which 
the purely bosonic gauge field $A_{\mu}(X)$ on the D $p$-brane world
volume couples to the boundary $\partial\Sigma$ of the string
world sheet $\Sigma$.
Our investigation is made in the rigid superspace
for simplicity.

\subsection{Variations of GS superstring action with boundary}
\label{sec:varGS}

The action of the type IIB GS superstring \cite{r:GS} is
\ba
&&S_{\rm GS}=-\int_{\Sigma}d^{2}\sigma\left[\sqrt{-g}
                  +{\cal L}_{\rm WZ}\right]
             -\oint_{\pS}dX^{\mu}A_{\mu}(X)~,\non
&&\qquad {\cal L}_{\rm WZ}
      =i\epsilon^{ab}\partial_{a}X^{\mu}
        (\bar{\theta}^{1}\Gamma_{\mu}\partial_{b}\theta^{1}
            -\bar{\theta}^{2}\Gamma_{\mu}\partial_{b}\theta^{2})
     {} -\epsilon^{ab}
         (\bar{\theta}^{1}\Gamma^{\mu}\partial_{a}\theta^{1})
         (\bar{\theta}^{2}\Gamma_{\mu}\partial_{b}\theta^{2})~.
\label{eq:gs}
\ea
where $\sigma^{a}$ $(a=0,1)$ are world sheet coordinates,
and $(X^{\mu}(\sigma),\theta^{A}_{\alpha}(\sigma))$
$(\mu =0,1,\ldots 9; \alpha =1,\ldots ,32; A=1,2)$ is the embedding
map from $\Sigma$ into the 1+9 dimensional superspace.
Here $\theta^{A}$ $(A=1,2)$ are Majorana-Weyl spinors
in 1+9 dimensional space-time with the same chiralities.
$g$ is the determinant of the induced metric $g_{ab}$
on $\Sigma$~:
\ba
&&g_{ab}=\Pi^{\mu}_{a}\Pi^{\nu}_{b}\eta_{\mu\nu},\non
&&\Pi^{\mu}_{a}=\partial_{a}X^{\mu}
    -i\bar{\theta}^{1}\Gamma^{\mu}\partial_{a}\theta^{1}
    -i\bar{\theta}^{2}\Gamma^{\mu}\partial_{a}\theta^{2}.
\ea

When the world sheet $\Sigma$ has no boundary,
the action (\ref{eq:gs}) is invariant under
${\cal N}=2$ space-time supersymmetry transformations,
\be
\delta_{\rm SUSY}\theta^{A}=\epsilon^{A},
\mbox{\hspace*{2em}}
  \delta_{\rm SUSY}X^{\mu}
    =-i\bar{\theta}^{1}\Gamma^{\mu}\delta_{\rm SUSY}\theta^{1}
     {}-i\bar{\theta}^{2}\Gamma^{\mu}\delta_{\rm SUSY}\theta^{2},
  \label{eq:strgs}
\ee
where $\epsilon^{A}$ are constant Majorana-Weyl spinors.
It is also invariant under local fermionic transformations
(so-called $\kappa$-symmetry transformations),
\be
\delta_{\kappa}\theta^{A}=\alpha^{A},
\mbox{\hspace*{2em}}
  \delta_{\kappa}X^{\mu}
     =i\bar{\theta}^{1}\Gamma^{\mu}\delta_{\kappa}\theta^{1}
      +i\bar{\theta}^{2}\Gamma^{\mu}\delta_{\kappa}\theta^{2},
 \label{eq:ktrgs}
\ee
where
\be
\alpha^{1}=(1+\tilde{\Gamma})\kappa^{1},
\mbox{\hspace*{2em}}
\alpha^{2}=(1-\tilde{\Gamma})\kappa^{2},
\mbox{\hspace*{2em}}
\tilde{\Gamma}=\frac{1}{2\sqrt{-g}}\Sigma_{\mu\nu}\Gamma^{\mu\nu},
\mbox{\hspace*{2em}}
\Sigma^{\mu\nu}=\epsilon^{ab}\Pi^{\mu}_{a}\Pi^{\nu}_{b}.
\label{eq:ka}
\ee
In the above,
$\kappa^{A}$ are Majorana-Weyl spinors depending on the
world sheet coordinates $\sigma^{a}$
and $\tilde{\Gamma}$ is subject to
$\tilde{\Gamma}^{2}=1$ and ${\rm Tr}(\tilde{\Gamma})=0$.

The lagrangian of the action (\ref{eq:gs}) is invariant
under these fermionic transformations modulo total derivative
terms.
In the presence of the world sheet boundary,
the variations of the action
(\ref{eq:gs}) under these fermionic transformations
leave the boundary terms.
The situation is parallel to that in the type IIA
theory \cite{r:EMM3}.
The explicit form of
$\delta_{\rm SUSY}S_{\rm GS}$ and $\delta_{\kappa}S_{\rm GS}$
is obtained
from eqs.(6) and (8) in ref.\cite{r:EMM3} by the replacement,
\be
\frac{1+\Gamma_{11}}{2}\theta\longmapsto\theta^{1}~,\qquad
\frac{1-\Gamma_{11}}{2}\theta\longmapsto\theta^{2}~.
\ee

In what follows we will determine the boundary conditions
along the line of the analysis in the type IIA theory \cite{r:EMM3}.
%
%
Because we consider the situation in which
the open superstring ends on the D $p$-branes,
%
%
we have
\be
\delta X^{\bmu}=0~, \qquad
\bar{\theta}^{1}\Gamma^{\bmu}\delta\theta^{1}
+\bar{\theta}^{2}\Gamma^{\bmu}\delta\theta^{2}=0,
\qquad \mbox{on $\partial\Sigma$}.\label{eq:dbcdbc}
\ee
%
%
Here $\umu (=0,1,\ldots,p)$ and $\bnu (=p+1,\ldots,9)$
denote, respectively, the directions
which are
parallel and perpendicular to the D $p$-brane.
We find that we are able to respect supersymmetry, $\kappa$-symmetry
and variational principle
by imposing, besides eq.(\ref{eq:dbcdbc}),
the boundary conditions,
\ba
&&(\bar{\theta^{1}}\Gamma_{\umu}\delta\theta^{1}
 {}-\bar{\theta^{2}}\Gamma_{\umu}\delta\theta^{2})
 {}-F_{\umu\unu}
   (\bar{\theta^{1}}\Gamma^{\unu}\delta\theta^{1}
  +\bar{\theta^{2}}\Gamma^{\unu}\delta\theta^{2})=0,
      \label{eq:nbcfrm}\\
&&\sqrt{-g}\,n^{a}\Pi_{a\umu}
  {}-F_{\umu\unu}\,n_{a}\epsilon^{ab}\Pi^{\unu}_{b}=0~,
\qquad\mbox{on $\pS$}~,
     \label{eq:nbcbos}
\ea
where $n^{a}$ denotes a unit vector normal to $\pS$,
and $F_{\mu\nu}=\partial_{\mu}A_{\nu}-\partial_{\nu}A_{\mu}$.


\subsection{Linear boundary conditions for fermions}
\label{sec:bclin}

We rewrite the boundary conditions
in a linear form w.r.t. fermions $\theta^{A}$ 
in order to preserve a fraction of supersymmetry.
We look for such boundary conditions first in the case
that $F_{\umu\unu}=0$.
We begin by focusing on eqs.(\ref{eq:dbcdbc}) and (\ref{eq:nbcfrm}).
We set the following ansatz:
\be
\theta^{A}_{\alpha}
=(~{{\Gp}_{\alpha}}^{\beta}\otimes M^{AB}~)\,\theta^{B}_{\beta}
\quad (\alpha,\beta =1,\ldots ,32; A,B=1,2)~,
\quad\mbox{on $\pS$}~,\label{eq:ans1}
\ee
where $\Gp \equiv \Gamma_{\underline{0}}\Gamma_{\underline{1}}
       \cdots \Gamma_{\underline{p}}$
and $M$ is a $2\times 2$ matrix.
In order that this ansatz is self-consistent,
$(\Gp\otimes M)\theta$ must be a Majorana-Weyl spinor with the
same chiralities as $\theta$ and
the relation $(\Gp\otimes M)^{2}=1$ should hold.
These consistency conditions require that 
$M^{*}=M$ and that $p$ should be odd.
Substituting eq.(\ref{eq:ans1}) into eqs.(\ref{eq:dbcdbc})
and (\ref{eq:nbcfrm}) with $F_{\umu\unu}=0$,
we find that
\be
M=\left\{
  \begin{array}{cl}
    \pm \sigma^{1}&\quad p=4m+1\\
    \pm i\sigma^{2}&\quad p=4m+3
  \end{array}\right. . \label{eq:sol}
\ee
It yields that, for any odd integer $p$,
\be
\theta^{1}=\pm \Gp\theta^{2}~,\quad\mbox{on $\pS$}~. \label{eq:any1}
\ee
By substituting eqs.(\ref{eq:ans1}) and (\ref{eq:sol})
into eq.(\ref{eq:nbcbos}),
we eventually find 
the desired boundary conditions:
\be
   \begin{array}{ll}
   \delta X^{\bmu}=0,& 
    \frac{1}{2}\left(1-\Gp\otimes M\right)\theta =0,\\
    n^{a}\partial_{a}X^{\umu}=0,&
   \frac{1}{2}\left(1+\Gp\otimes M\right)n^{a}\partial_{a}\theta =0~.
   \end{array}
  \label{eq:bczerof}
\ee

Let us now turn on the field strength $F_{\umu\unu}$
on the D $p$-brane world volume.
We restrict ourselves to the constant
$F_{\umu\unu}$.
By an analysis similar to that in 
the type IIA theory \cite{r:EMM3},
we obtain
the linear boundary conditions
that reproduce the conditions (\ref{eq:dbcdbc}) and (\ref{eq:nbcfrm}),
\be
\theta=
 e^{-\frac{1}{2}Y_{\umu\unu}(\Gamma^{\umu\unu}\otimes\sigma^{3})}
  (\Gp\otimes M)\theta~,\quad\mbox{on $\pS$}~,\label{eq:ymunu}
\ee
where $Y_{\umu\unu}$ is defined such that
$F_{\umu\unu}=(\tanh Y)_{\umu\unu}$.
In general, however, it is difficult to
rewrite the condition (\ref{eq:nbcbos}) in a simple form,
since it is fairly non-linear.

We point out that
the result (\ref{eq:ymunu}) is consistent with that obtained in the
light-cone gauge~\cite{r:GG}
and yields the same supersymmetry breaking pattern
as is given in ref.\cite{r:bkop}.

\subsection{Analytic continuation of $\theta^{2}$}
When IKKT showed the correspondence between their matrix model
and the type IIB theory in the Schild gauge,
they started from the action in which 
the signs of the $\theta^{2}$-bilinear terms are 
reversed:
\ba
&&S_{\rm GS}^{\rm (IKKT)}=-\int_{\Sigma} d^{2}\sigma\left[
  \sqrt{-\frac{1}{2}\Sigma^{2}}+i\epsilon^{ab}\partial_{a}X^{\mu}
   (\bar{\theta}^{1}\Gamma_{\mu}\partial_{b}\theta^{1}
   +\bar{\theta}^{2}\Gamma_{\mu}\partial_{b}\theta^{2})
  \right.\non
&&\mbox{\hspace*{12em}}\left.
+\epsilon^{ab}
     (\bar{\theta}^{1}\Gamma^{\mu}\partial_{a}\theta^{1})
     (\bar{\theta}^{2}\Gamma_{\mu}\partial_{b}\theta^{2})
   \right], \label{eq:ikkt}
\ea
where $\Sigma^{\mu\nu}$ is defined in eq.(\ref{eq:ka})
and we note that $\frac{1}{2}\Sigma^{2}=g$.
In the present case, however, $\Pi_{a}^{\mu}$ is defined as
\be
\Pi_{a}^{\mu}=\partial_{a}X^{\mu}
   -i\bar{\theta}^{1}\Gamma^{\mu}\partial_{a}\theta^{1}
   +i\bar{\theta}^{2}\Gamma^{\mu}\partial_{a}\theta^{2}.
\ee

In order to obtain the action (\ref{eq:ikkt}) from the conventional
one,
we need to perform an analytic continuation,
$\theta^{2}\rightarrow i\theta^{2}$ \cite{r:ikkt}.
Here we should note that the Dirac conjugation
$\overline{\psi}$ of a spinor $\psi$
is now supposed to be defined by eq.(\ref{eq:dirac}):
$\overline{\psi}=-\psi^{T}{\cal C}^{-1}$.
Hereafter we use this redefinition
for all the spinors regardless of whether they are Majorana or not.


The ${\cal N}=2$ space-time supersymmetry and the $\kappa$-symmetry
transformations are given by reversing the
signs of $\theta^{2}$-bilinear terms in eqs.(\ref{eq:strgs}) and
(\ref{eq:ktrgs}).


In what follows,  we consider the boundary conditions for
a type IIB Dirichlet open string defined by the action
(\ref{eq:ikkt}).
We now concentrate on the case in which the field strength
$F_{\mu\nu}$ on the D $p$-branes is switched off.
The boundary conditions
are obtained by reversing the sign of the $\theta^{2}$-bilinear terms
in eqs.(\ref{eq:dbcdbc}), (\ref{eq:nbcfrm}) and (\ref{eq:nbcbos})
with $F_{\mu\nu}=0$.
We should rewrite them into a linear form
w.r.t. fermionic coordinates $\theta^{A}$.
Applying to eq.(\ref{eq:any1})
the analytic continuation,
$\theta^{2}\rightarrow i\theta^{2}$,
we obtain the linear boundary conditions,
\be
\theta^{1}=\pm i \Gamma_{(p)}\theta^{2}~,
\qquad \mbox{on $\pS$}~.\label{eq:any2}
\ee
It implies that $M$ in eq.(\ref{eq:ans1}) is replaced by
\be
M=\left\{
  \begin{array}{cl}
    \mp\sigma^{2}&\quad p=4m+1\\
    \pm i\sigma^{1}&\quad p=4m+3
  \end{array}
   \right. .
\ee  
We remark on the Majorana condition for $\theta$.
{}From the above we find that $M^{*}\neq M$.
It means that these linear boundary conditions
do not satisfy the Majorana condition.
In fact there is not such a matrix $M$
as fulfill the boundary conditions and the Majorana condition
simultaneously.
In the present case, however, there are no strong reasons
to impose the Majorana condition on $\theta$.
This is because we have performed the analytic continuation
and the Majorana condition become subtle.
We will henceforth ignore the Majorana condition.


\section{Matrix Regularization of an Open Superstring}        
\label{sec:mros}                                              
In this section, we regularize a type IIB Dirichlet
open superstring by a matrix model,
by using prescriptions similar to those in ref.\cite{r:ikkt}.
We restrict ourselves to the case that the open string
ends on two parallel D $p$-branes or on a single D $p$-brane.
In this situation, only the DD and NN sectors emerge and we do
not have to consider DN or ND sector.

\subsection{Schild gauge formulation}              
\label{sec:sgf}                                    
In this section we consider an open superstring in the
Schild gauge,
\be
\psi\equiv\theta^{1}=\theta^{2}.\label{eq:schild}
\ee
As is done by IKKT, we introduce an auxiliary
field $\sqrt{h}$,
which is a positive definite scalar density on the
world sheet. The action is rewritten \cite{r:ikkt} into
\be
S_{\rm Schild}=\int_{\Sigma} d^{2}\sigma\left[
  \sqrt{h}~\alpha\left(\frac{1}{4}\{X^{\mu},X^{\nu}\}^{2}
  {}-\frac{i}{2}\overline{\psi}\Gamma^{\mu}
      \{X_{\mu}\, ,\, \psi\}\right)+\beta\sqrt{h}~\right],
 \label{eq:schtype}
\ee
where $\{ *,*\}$ is the Lie bracket defined as
$\{ X,Y\}=\frac{\epsilon^{ab}}{\sqrt{h}}~
\partial_{a}X\,\partial_{b}Y$,
for arbitrary functions  $X(\sigma)$ and $Y(\sigma)$
on the world sheet.
Combining with the $\kappa$-symmetry transformations,
we can define the Schild gauge-preserving supersymmetry
transformations \cite{r:ikkt} as
\be
\left\{
 \begin{array}{ccl}
  \delta^{(1)}\psi&=&
   -\frac{1}{2\sqrt{h}}\sigma_{\mu\nu}\Gamma^{\mu\nu}\eta\\
  \delta^{(1)}X^{\mu}&=&i\overline{\eta}\Gamma^{\mu}\psi\\
  \delta^{(1)}\sqrt{h}&=&0
 \end{array}
\right. ,\qquad
\left\{
 \begin{array}{ccl}
  \delta^{(2)}\psi &=&\xi\\ 
  \delta^{(2)}X^{\mu}&=&0\\
  \delta^{(2)}\sqrt{h}&=& 0
 \end{array}
\right. .\label{eq:sush}
\ee

The boundary conditions are modified to
\ba
\delta X^{\bmu}=0~,&& \psi=\pm i\Gp\psi~,\non
n^{a}\partial_{a}X^{\umu}=0~,&&
n^{a}\partial_{a}\psi=\mp i\Gp n^{a}\partial_{a}\psi~,
\quad\mbox{on $\pS$}~.
  \label{eq:bcsch}
\ea
In order that the above conditions for the fermionic sector should
be self-consistent,
$\Gp$ have to satisfy
$(i\Gp)^{2}=1$, i.e. ${\Gp}^{2}=-1$.
Combined with eq.(\ref{eq:gp}), this implies that
\be
p=4m+3=3,\ 7~. \label{eq:allowed}
\ee
We find that the Schild-type gauge choice restricts
the possible dimensions of D branes.
Such a restriction must be a gauge artifact.
The situation seems to be similar to the difficulties
in describing transverse five-branes in the BFSS matrix theory.
In the following we restrict ourselves to the case that $p=4m+3$.

We make some comments on the gauge symmetry on the open
world sheet.
When the world sheet $\Sigma$ is a closed surface,
the Schild-type action (\ref{eq:schtype}) has 
gauge symmetry whose gauge group is
area-preserving diffeomorphisms (APD) on $\Sigma$ :
\be
\delta_{\rm gauge}X^{\mu}(\sigma)
=-\{\zeta(\sigma)\, ,X^{\mu}(\sigma)\}~,
\quad \delta_{\rm gauge}\psi(\sigma)
=-\{\zeta(\sigma)\, ,\psi(\sigma)\}~,\label{eq:apdgauge}
\ee
where $\zeta(\sigma)$ is
an infinitesimal arbitrary function globally well-defined on $\Sigma$.
In order that the boundary contributions of
$\delta_{\rm gauge}S_{\rm Schild}$ should vanish,
the transformation parameter $\zeta (\sigma)$ has to obey
the Dirichlet boundary condition.
The situation is the same as the open supermembrane in the light-cone
gauge \cite{r:EMM3}.

\subsection{Unbroken supersymmetry algebra}                  %
  \label{sec:ubsusy}                                         %
Let us now consider the commutator algebra of the
unbroken supersymmetry transformations.
Because they
have to preserve the boundary conditions (\ref{eq:bcsch}),
the parameters in eq.(\ref{eq:sush}) are subject to
the chirality projection on the D $p$-brane
world volume: $\eta=\mp i\Gp\eta~$ and
$\xi=\pm i\Gp\xi$.
The resulting commutator algebra turns out to be almost the same
as that in ref.\cite{r:ikkt}. The only difference resides in
the variation of the bosonic coordinates,
\be
{}[\delta^{(1)}_{\eta}\, ,\,\delta^{(2)}_{\xi}]X_{\umu}
=-i\overline{\eta}\,\Gamma_{\umu}\,\xi, \qquad
{}[\delta^{(1)}_{\eta}\, ,\,\delta^{(2)}_{\xi}]X_{\bmu}=0.
\ee
This implies that the super-translation algebra reduces
to that on the D $p$-brane world volume.

\subsection{Matrix regularization of open superstring} 
In this section we investigate the matrix model
for a Dirichlet open superstring in the Schild gauge.
We consider only the case in which the topology of the
world sheet $\Sigma$ is a cylinder.
We identify the world sheet coordinate such that 
$\tau(=\sigma^{0})$ parametrizes the $S^{1}$-direction
(i.e. $\tau\sim \tau+1$) and $\sigma(=\sigma^{1})\in [0,\frac{1}{2}]$
parametrizes the $I$-direction of the cylinder $S^{1}\times I$.

We introduce the notations, $\psi=\psi^{(D)}+\psi^{(N)}$,
where
\be
\psi^{(D)}\equiv \frac{1}{2}(1\mp i\Gp)\psi~,
\quad \psi^{(N)}\equiv\frac{1}{2}(1\pm i\Gp)\psi.
\ee
{}From eq.(\ref{eq:bcsch}) we find that
$\psi^{(D)}$ and $\psi^{(N)}$ obey the Dirichlet and
the Neumann boundary conditions respectively.

We approximate the real fields on the world sheet by
$N\times N$ hermitian matrices.
By using the same reasoning that is made in
ref.\cite{r:EMM3},
we obtain the correspondence rules:
\ba
\mbox{DD sector\,: $X^{\bmu}$~, $\psi^{(D)}$}
&\stackrel{N\rightarrow\infty}{\longleftarrow}&
\mbox{$N\times N$ antisymmetric matrices}~,\non
\mbox{NN sector\,: $X^{\umu}$~, $\psi^{(N)}$}
&\stackrel{N\rightarrow\infty}{\longleftarrow}&
\mbox{$N \times N$ symmetric matrices}~.
\ea
As is mentioned in sec.\ref{sec:sgf},
the transformation parameters of 
APD must belong to the Dirichlet sector. It follows that
\be
\mbox{parameters of APD}
\ \stackrel{N\rightarrow\infty}{\longleftarrow}
\ \mbox{$N\times N$ antisymmetric matrices}.\label{eq:paraapd}
\ee

We can now write down the matrix model to regularize an
open superstring.
Following the standard procedure, we replace
real fields on the world sheet,
$\int_{\Sigma}d^{2}\sigma\,\sqrt{h}$ and $i\{*\,,\,*\}$
with
$N\times N$ hermitian matrices, Tr and $[*\,,\,*]$
respectively.
Consequently, we obtain the matrix regularization
of the action (\ref{eq:schtype}):
\ba
&&S=\alpha\left[-\frac{1}{4}{\rm Tr}
 \left(\,[X^{\umu}\,,\,X^{\unu}]^{2}+2[X^{\umu}\,,\,X^{\bnu}]^{2}
        +[X^{\bmu}\,,\,X^{\bnu}]^{2}\,\right)\right.\non
&&\mbox{\hspace*{3.5em}}
 {}-\frac{1}{2}{\rm Tr}
 \left(\, 
 + \overline{\psi}^{(D)}\Gamma^{\bmu}[X_{\bmu}\,,\,\psi^{(D)}]
 + \overline{\psi}^{(N)}\Gamma^{\bmu}[X_{\bmu}\,,\,\psi^{(N)}]
\right.\non
&&\mbox{\hspace*{7.5em}}\left.\left.
 + \overline{\psi}^{(D)}\Gamma^{\umu}[X_{\umu}\,,\,\psi^{(N)}]
 +\overline{\psi}^{(N)}\Gamma^{\umu}[X_{\umu}\,,\,\psi^{(D)}]
 \,\right)\right]+\beta\,{\rm Tr}~{\bf 1}~.\label{eq:maopen}
\ea
The APD gauge transformation (\ref{eq:apdgauge}) becomes
\be
\delta_{\rm gauge}X^{\mu}=i[\zeta\,,\,X^{\mu}]~,
\quad
\delta_{\rm gauge}\psi=i[\zeta\,,\,\psi]~,
\ee
where $\zeta$ is an antisymmetric matrix as is found in
eq.(\ref{eq:paraapd}).
This gauge transformation can be identified with
$SO(N)$ gauge transformation. 
$X^{\bmu}$ and $\psi^{(D)}$ belong to the adjoint representation
and $X^{\umu}$ and $\psi^{(N)}$ belong to the
2nd rank symmetric representation of $SO(N)$ respectively.
The situation is very similar to the case of a light-cone
open supermembrane~\cite{r:EMM3}\cite{r:kr}.

We point out here that in the $p=3$ case
the bosonic and the fermionic physical degrees of freedom
match.
The matter contents in this case are given as follows,
\begin{eqnarray}
\mbox{Neumann sector}&:&\left\{
 \begin{array}{ll}
   \mbox{bosonic}& X^{\underline{0}},\ X^{\underline{1}},
     \ X^{\underline{2}},\ X^{\underline{3}}\\
   \mbox{fermionic}& \psi^{(N)}\equiv \frac{1\pm i\Gp}{2}\psi
 \end{array}
\right.\quad ,\non
\mbox{Dirichlet sector}&:&\left\{
 \begin{array}{ll}
  \mbox{bosonic}& X^{\overline{4}},
     \ X^{\overline{5}},\ X^{\overline{6}},
     \ X^{\overline{7}},\ X^{\overline{8}},
     \ X^{\overline{9}}\\
  \mbox{fermionic}& \psi^{(D)}\equiv \frac{1\mp i\Gp}{2}\psi
 \end{array}
\right. \quad.
\end{eqnarray}
It is obvious that, in the Neumann sector,
the physical degrees of freedom of bosons and fermions
are both four.
In the Dirichlet sector, while we have four fermionic
degrees of freedom, there appear to be six bosonic ones.
Owing to the gauge symmetry, however, the
number of the latter reduces by two.
Thus the bosonic and fermionic physical degrees of freedom
match in each sector.

We note that the above matter contents
can be interpreted as the zero volume limit
of the six dimensional ${\cal N}=1$
$SO(N)$ super Yang-Mills theory which couples to
a hypermultiplet in the second rank
symmetric representation of the gauge group.
We hope that this model will play a role in understanding
D 3-branes, which are sometimes difficult to analyze
because of their self-duality.

Finally we mention the relationship
between the recently proposed large-$k$ $USp(2k)$
matrix model \cite{r:it} and our $SO(N)$ model.
The authors of \cite{r:it} find that, in order to
have unbroken supersymmetry,
there are two ways of
projecting the ten bosonic matrix coordinates into
$n_{-}$ components in the adjoint representation of $USp(2k)$
and $n_{+}$ ones in the antisymmetric representation.
One is $(n_{-},n_{+})=(6,4)$ and the other is
$(n_{-},n_{+})=(2,8)$.
They correspond to (3+1)-
and (7+1)-dimensional orientifold fixed planes respectively.
This result should be naturally derived by
extending our analysis to an unoriented superstring.
What we have learned are that $SO(N)$ matrix models and $USp(2k)$ ones
incorporate D branes and orientifold planes respectively,
and that the possible dimensions of these objects undergo the
identical restriction.
It might be interesting to further investigate
these relationship especially in the type I theory.

\vskip 5mm

\noindent{\bf Acknowledgements:} \hskip 10mm
We would like to thank M. Ninomiya, M. Fukuma, S. Hirano,
H. Kunitomo and K. Sugiyama for invaluable discussions,
comments and encouragement.

\section*{Appendix}  \label{sec:conv}                    %
1+9 dimensional gamma matrices $\Gamma^{\mu}$
$({\mu}=0,1,\ldots ,9)$ satisfy the $SO(1,9)$ Clifford algebra
$\{\Gamma^{\mu},\Gamma^{\nu}\}=-2\eta^{\mu\nu}$,
where we use the almost plus sign convention as the ten dimensional
Minkowski metric $\eta_{\mu\nu}=\mbox{diag}(-,+,\ldots ,+)$ .
In this convention $\Gamma^{0}$ is defined to be hermitian and
$\Gamma^{i}$ $(i=1,\ldots ,9)$ are to be anti-hermitian.
Thus we find
$\Gamma_{\mu}^{\dagger}=\Gamma^{0}\Gamma_{\mu}\Gamma^{0}$.
The charge conjugation matrix ${\cal C}$ satisfies
$\Gamma_{\mu}^{T}=-{\cal C}^{-1}{\Gamma_{\mu}}{\cal C}$,
and ${\cal C}^{T}=-{\cal C}$.
It follows that
\be
{\Gp}^{2}=(-)^{\frac{1}{2}p(p+3)}I_{32}~,\qquad
{\Gp}^{T}=(-)^{\frac{1}{2}(p+1)(p+2)}{\cal C}^{-1}\Gp{\cal C}~,
\label{eq:gp}
\ee
where $\Gp \equiv \Gamma_{01\cdots p}$.
The charge conjugation of a spinor $\theta$ is defined as
$\theta^{c}={\cal C}\bar{\theta}^{T}$,
where $\bar{\theta}\equiv\theta^{\dagger}\Gamma^{0}$ is the
Dirac conjugate of $\theta$.
It follows that the Majorana condition $\psi =\psi^{c}$
means that
\be
\overline{\psi}=-\psi^{T}{\cal C}^{-1}
\qquad\mbox{for Majorana spinor $\forall\psi$}~.\label{eq:dirac}
\ee



\begin{thebibliography}{99}
\bibitem{r:bfss}T. Banks, W. Fischler, S.H. Shenker and
   L. Susskind, Phys. Rev. {\bf D55} (1997) 5112.
\bibitem{r:ikkt}N. Ishibashi, H. Kawai, Y. Kitazawa and A. Tsuchiya,
  Nucl. Phys. {\bf B498} (1997) 467.\\
  M. Fukuma, H. Kawai, Y. Kitazawa and A. Tsuchiya,
  {\it ``String Field Theory from IIB Matrix Model''},
    hep-th/9705128.\\
  H. Aoki, S. Iso, H. Kawai, Y. Kitazawa and T. Tada,
  {\it ``Space-Time Structures from IIB Matrix Model''},
    hep-th/9802085.
\bibitem{r:bps}
T. Banks, N. Seiberg and S. Shenker,
  Nucl. Phys. {\bf B490} (1997) 91.\\
I. Chepelev, Y. Makeenko and K. Zarembo, Phys. Lett. {\bf B400}
  (1997) 43.\\
A. Fayyazuddin and D.J. Smith,
  Mod. Phys. Lett. {\bf A12} (1997) 1447
\bibitem{r:EMM3}K. Ezawa, Y. Matsuo and K. Murakami,
   {\it ``Matrix Regularization of Open Supermembrane
     --towards M-theory five-brane via open supermembrane''},
     hep-th/9707200.
\bibitem{r:dWHN}B. de Wit, J. Hoppe and H. Nicolai,
  Nucl. Phys. {\bf B305} (1988) 545.\\
    B. de Wit, K. Peeters and J. Plefka,
    {\it ``Open and Closed Supermembranes with Winding''},
    hep-th/9710215.
\bibitem{r:GG}M.B. Green and M. Gutperle,
   Nucl. Phys. {\bf B476} (1996) 484.\\
  see also
    C. Chu, P. Howe and E. Sezgin,
   {\it ``Strings and D-Branes with Boundaries''},
   hep-th/9801202.
\bibitem{r:GS}M.B. Green and J.H. Schwarz,
   Phys. Lett. {\bf B136} (1984) 367;
   Nucl. Phys. {\bf B143} (1984) 285.
\bibitem{r:bkop}E. Bergshoeff, R. Kallosh, T. Ort\'{\i}n
   and G. Papadopoulos, Nucl. Phys. {\bf B502} (1997) 149.
\bibitem{r:kr}N. Kim and S.-J. Rey,
  Nucl. Phys. {\bf B504} (1997) 187.
\bibitem{r:it}H. Itoyama and A. Tokura,
   {\it ``$USp(2k)$ Matrix Model~: Nonperturbative Approach to
       Orientifolds''}, hep-th/9801084.

\end{thebibliography}
\end{document}